# Towards large eddy simulations of premixed turbulent flames without a combustion model

Andrei N. Lipatnikov

*Department of Mechanics and Maritime Sciences, Chalmers University of Technology, Gothenburg SE-412 96, Sweden*

______________________________________________________________________

**Abstract**
The paper aims at assessing a hypothesis that resolution required to evaluate fuel consumption and heat release rates by directly (i.e., without a subgrid model of unresolved influence of small-scale turbulent eddies on the local flame) processing filtered fields of density, temperature, and species mass fractions should be significantly finer than resolution required to directly compute flame surface density by processing the same filtered fields. For this purpose, box filters of various widths $\Delta$ are applied to three-dimensional Direct Numerical Simulation data obtained earlier by Dave et al. (Combust. Flame 196 (2018) 386-399) from a statistically one-dimensional and planar, moderately lean $H_2$/air complex-chemistry flame propagating in a box under conditions of sufficiently intense small-scale turbulence (Karlovitz number is larger than unity and a ratio of laminar flame thickness $\delta_L$ to Kolmogorov length scale is about 20). Results confirm this hypothesis and show that the mean flame surface density and area can be predicted with acceptable accuracy by processing filtered combustion progress variable fields computed using a sufficiently wide filter, e.g., $\Delta/\delta_L = 4/3$. Such an approach does not require a model of the influence of subgrid turbulent eddies on flame surface density provided that $\Delta$ and $\delta_L$ are of the same order of magnitude. Good performance of this approach is attributed to inability of small-scale (when compared to $\delta_L$) turbulent eddies to substantially change the local flame structure, which, nevertheless, is significantly perturbed by larger turbulent eddies that strain the local flame.

*Keywords:* premixed turbulent combustion; large eddy simulation; flame surface density; modeling; numerical resolution

## 1. Introduction

The problem of evaluation of mean reaction rates in turbulent flames stems from a highly non-linear dependence of rates of many important reactions on temperature and has been challenging the combustion community over decades. Earlier, a number of models was developed to compute mean reaction rates within the framework of Reynolds-Averaged Navier-Stokes (RANS) approach, as reviewed elsewhere [1,2]. Today, the focus of Computational Fluid Dynamics (CFD) studies of turbulent burning is shifted to Large Eddy Simulations (LES) [3-6], which deal with quantities filtered over sufficiently small volumes, thus, allowing researchers to explore flame dynamics by directly resolving local processes in a wide range of turbulence spectrum. However, scales associated with the influence of the smallest turbulent eddies on the local flame are rarely resolved and such subgrid effects still require modeling. For this purpose, both LES counterparts of RANS models and LES-specific models were developed. The former group involves, e.g., Flame Surface Density (FSD) or flame wrinkling models [7-10], scalar dissipation rate models [9-11], presumed PDF models [12,13], transported PDF approach [14,15], etc. Thickened flame models [16,17] and dynamic methods [18-20] are well-known members of the latter group.

Rapid progress in computer hardware and software has continuously been extending the range of scales resolvable in LES of turbulent burning, thus, enabling a resolution of several computational cells per laminar flame thickness in simulations of laboratory measurements. Accordingly, there is a growing body of LES studies [21-32] that do not invoke any model of the influence of small-scale turbulence on a flame (henceforth, "a combustion model" for brevity), but directly evaluate filtered reaction rates using the locally filtered values of density, temperature, and species mass fractions. Recent *a posteriori* studies supported this approach, provided that a ratio of filter width $\Delta$ to laminar flame thickness $\delta_L$ is sufficiently small, e.g., 0.25 [31], 0.2 [23], 0.125 [30], or even 0.05 [32]. On the contrary, LES results [28] computed using different combustion models show that, if $\Delta \cong \delta_L$, the discussed simplest approach performs substantially worse than other models assessed in the cited paper. Results of *a priori* studies [33-36], obtained by filtering Direct Numerical Simulation (DNS) data, also show that numerical resolution required by this approach is significantly less than laminar flames thickness, e.g., $\Delta = 0.2\delta_L$ [35]. However, so fine resolution does not seem to be feasible in applied CFD research into turbulent combustion under elevated pressures and elevated temperatures in engines, where the thickness $\delta_L$ is very small [36].

Nevertheless, there seems to be another way to running LES of turbulent burning without any combustion model. First, starting from the pioneering work by Damköhler [37], an increase in burning velocity $U_t$ by turbulence is often attributed mainly to an increase $\delta A$ in flame surface area stretched by turbulent eddies. Recent DNS studies [38-41] do show that $U_t \cong S_L \delta A$ even in moderately lean H$_2$/air turbulent flames [39], see also Fig. 1 in Sect. 2, or in equidiffusive highly turbulent flames [40], where local flame speed can differ significantly from the laminar flame speed $S_L$ and be even negative [39,40]. In very lean hydrogen/air turbulent flames, $U_t$ can be significantly larger than $S_L \delta A$ [40] due to differential diffusion effects reviewed elsewhere [42]. Since there is no widely accepted model for predicting an increase in $U_t/S_L$ due to such effects, they are beyond the scope of this work, whose focus is solely placed on comparing different approaches to LES of premixed equidiffusive flames, where $U_t \cong S_L \delta A$.

Second, various experimental and numerical data indicate that flame surface area is weakly affected by small-scale (when compared to $\delta_L$) turbulent eddies. For instance, measurements and computations of interaction of a laminar flame with a single vortex or a vortex pair show that too small vortices decay rapidly and do not substantially perturb the flame, see review articles [43,44] and recent papers [45,46]. Moreover, small-scale turbulent eddies may weakly affect a premixed flame, because residence time of the eddies within the flame is significantly reduced due to combustion-induced acceleration of the local flow in the flame-normal direction [47,48]. Accordingly, measurements of fractal characteristics of flame surfaces in turbulent flows, reviewed elsewhere [49], show that inner cut-off scale of the surface wrinkles is significantly (by a factor of three or more) larger than $\delta_L$. Recent DNS data [50-52] also indicate weak influence of small-scale turbulent eddies on a flame surface. For instance, contribution of eddies smaller than $2\delta_L$ to the total tangential strain rate of flame surface was reported to be as low as 10% [51,52].

Therefore, exploring the following hypothesis appears to be of interest: If LES aims at computing filtered flame surface area $\delta A$, rather than filtered reaction rates, then, the use of a moderately fine mesh (e.g., $\Delta \cong \delta_L$) could allow researchers to directly (i.e., without a flame surface density or flame wrinkling model [7-10]) obtain the area by processing filtered scalar fields and, subsequently, evaluate turbulent burning velocity $U_t \cong S_L \delta A$. The present work aims at (i) assessing this simple hypothesis and (ii) comparing resolution requirements associated with direct evaluation of filtered surface area and filtered reaction rates.

For these purposes, DNS database created by Dave et al. [53,54] and analyzed also by the present author [36,48,55-58] will be used. The DNS attributes and applied diagnostic tools are briefly described in Sect. 2. Results are reported and discussed in Sect. 3, followed by conclusions.

## 2. DNS attributes and diagnostic methods

Since the DNS attributes are reported elsewhere [53-58], only their summary is given below.

A statistically planar and one-dimensional, lean hydrogen-air turbulent flame propagating in a cuboid ($19.18 \times 4.8 \times 4.8$ mm) was studied by adopting the Pencil code [59] to numerically solve unsteady and three-dimensional continuity, compressible Navier-Stokes, species and energy transport equations supplemented with the mixture-averaged molecular transfer model and a detailed chemical mechanism (9 species and 21 reactions) by Li et al. [60]. The cuboid was meshed with a uniform grid of $960 \times 240 \times 240$ cells. At the transverse sides, boundary conditions were periodic. At the inlet and outlet, Navier-Stokes characteristic boundary conditions [61] were set.

To pre-generate homogeneous isotropic turbulence in another cube with the periodic boundary conditions, large-scale forcing was adopted [53]. The turbulence was allowed to evolve until a statistically stationary state was reached. At this final stage [53], the rms velocity $u' = 6.7$ m/s; an integral length scale $L = 3.1$ mm; an integral time scale $\tau_t = L/u' = 0.46$ ms; turbulent Reynolds number $Re_t = u'L/\nu = 950$; Kolmogorov time scale $\tau_\eta = (\nu/\langle \varepsilon \rangle)^{1/2} = 0.015$ ms; and Kolmogorov length scale $\eta = (\nu^3/\langle \varepsilon \rangle)^{1/4} = 0.018$ mm. Here, $\langle \varepsilon \rangle = \langle 2\nu S_{ij} S_{ij} \rangle$ is the turbulence dissipation rate averaged over the cube; $\nu$ is the gas kinematic viscosity; $S_{ij} = (\partial u_i/\partial x_j + \partial u_j/\partial x_i)/2$ is the rate-of-strain tensor; and the summation convention applies to repeated indexes.

At $t = 0$, a pre-computed planar laminar flame (the equivalence ratio $\Phi = 0.81$, pressure $P = 1$ bar, and unburned gas temperature $T_u = 310$ K) was embedded into the cuboid at $x = x_0$. The laminar flame speed $S_L$, thickness $\delta_L^T = (T_b - T_u)/\max|\nabla T|$, and time scale $\tau_f = \delta_L^T/S_L$ are equal to 1.84 m/s, 0.36 mm, and 0.20 ms, respectively. Here, subscripts $u$ and $b$ designate unburned and burned gases, respectively. Subsequently, the flame was wrinkled and stretched by the pre-generated turbulence, which was continuously injected into the computational domain through its left boundary $x = 0$ and decayed along the $x$-direction.

The Karlovitz number $Ka = \tau_f/\tau_\eta$ and the Damköhler number $Da = \tau_t/\tau_f$, evaluated using characteristics of the pre-generated turbulence, are equal to 13 and 2.35, respectively. Due to decay of the injected statistically stationary turbulence with distance $x$ from the inlet, the turbulence characteristics averaged over the cuboid cross-section nearest to a plane where $\bar{c}_T(x) = 0.01$ (leading edge of the mean flame brush) are different [48]: $u' = 3.3$ m/s; Taylor length scale $\lambda = \sqrt{10\nu_u \bar{k}/\bar{\varepsilon}} = 0.25$ mm or $0.69\delta_L^T$; $\eta = 0.018$ mm or $0.05\delta_L^T$; $\tau_\eta = 0.087$ ms; $Re_\lambda = u'\lambda/\nu_u = 55$; and $Ka = 2.3$ is much less than $(\delta_L/\eta)^2 \cong 400$, because $S_L \delta_L/\nu_u \gg 1$ in moderately lean hydrogen-air mixtures [62]. Here, $k = u_j' u_j'/2$ is turbulent kinetic energy; $u_j' = u_j - \bar{u}_j$ designates $j$-th component of fluctuating velocity vector; $c_T = (T - T_u)/(T_b - T_u)$ is a temperature-based combustion progress variable; and overbars refer to time- and transverse-averaged quantities sampled at 55 instants from 1.29 to 1.57 ms.

The focus of the present analysis is placed on the influence of the width $\Delta$ of a box filter applied to the DNS data on (i) generalized flame surface density $|\nabla \widehat{c_F}|$ or $|\nabla \widehat{c_T}|$ and (ii) fuel consumption rate $\widehat{\dot{\omega}_F} \equiv \dot{\omega}_F(\hat{\rho}, \hat{T}, \hat{Y}_k)$ or heat release rate $\widehat{\dot{\omega}_T} \equiv \dot{\omega}_T(\hat{\rho}, \hat{T}, \hat{Y}_k)$ evaluated using filtered density $\hat{\rho}(\mathbf{x}, t)$, temperature $\hat{T}(\mathbf{x}, t)$, and species mass fraction $\hat{Y}_k(\mathbf{x}, t)$. Here, $c_F = 1 - Y_F/Y_{F,u}$ is a fuel-based combustion progress variable, $\hat{q}(\mathbf{x}, t)$ designates filtered value of the quantity $q(\mathbf{x}, t)$, and the filter width $\Delta$ is varied from $0.11\delta_L$ to $4.44\delta_L$.

Reported for these purposes are spatial variations of time- and transverse-averaged values $\bar{\hat{q}}(\bar{c}_F)$ of the aforementioned filtered quantities within the mean flame brush, with the $x$-dependence of $\bar{\hat{q}}(x)$ being converted to its $\bar{c}_F$-dependence using the monotonous profile $\bar{c}_F(x)$ of time- and transverse-averaged fuel-based combustion progress variable. Moreover, the following integrals:

$$\delta A_T(t) = \frac{1}{A_0} \iiint |\nabla \widehat{c_T}|(\mathbf{x}, t) d\mathbf{x}, \tag{1}$$

$$\delta A_F(t) = \frac{1}{A_0} \iiint |\nabla \widehat{c_F}|(\mathbf{x}, t) d\mathbf{x}, \tag{2}$$

which characterize filtered flame surface areas for various $\Delta$, will be compared with normalized turbulent burning velocities evaluated by integrating the raw DNS data over the computational domain, i.e.,

$$\frac{U_t^T(t)}{S_L} = \frac{1}{\rho_u S_L (T_b - T_u) A_0} \iiint \dot{\omega}_T(\mathbf{x}, t) d\mathbf{x}, \tag{3}$$

$$\frac{U_t^F(t)}{S_L} = -\frac{W_F}{\rho_u S_L Y_{F,u} A_0} \iiint \dot{\omega}_F(\mathbf{x},t) d\mathbf{x}. \quad (4)$$

Here, $A_0$ is the cuboid cross-section area, $W_F$ is fuel molecular weight, subscript and superscript F or T refers to fuel-based (fuel mass fraction and consumption rate) or temperature-based (temperature and heat release rate) framework. When discussing trends observed in both frameworks, subscripts or superscripts F and H will be omitted in the following.

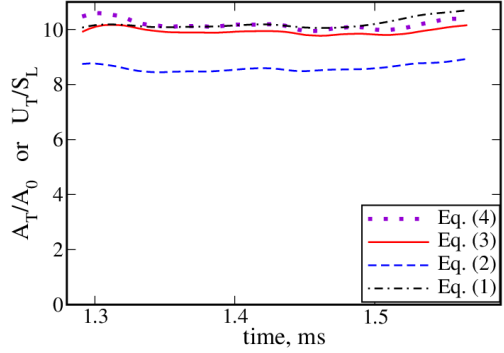

Fig. 1. Evolution of normalized turbulent burning velocities $U_t/S_L$ (violet dotted and red solid lines) and generalized flame surface area $A_0^{-1} \iiint |\nabla c| d\mathbf{x}$ (blue dashed and black dotted-dashed lines) sampled from the DNS data within fuel (violet dotted and blue dashed lines) or temperature (red solid and black dotted-dashed lines) framework.

The direct comparison of $\delta A(t)$ and $U_t(t)/S_L$ is justified by the fact that, for moderately lean H$_2$-air mixtures like the mixture used in the present study, the influence of differential diffusion phenomena on a mean turbulent burning rate is sufficiently weak [39-41,54,63-65]. For instance, Fig. 1 shows that $U_t(t)/S_L$ is close to the area increase $\delta A(t)$ sampled directly from the DNS data, i.e., calculated by substituting $|\nabla \hat{c}|(\mathbf{x},t)$ with $|\nabla c|(\mathbf{x},t)$ in Eq. (1) or (2). This trend is well pronounced within the temperature framework, i.e., when $U_t^T(t)/S_L$ is compared with $\delta A_T(t)$. Differences in $U_t^F(t)/S_L$ and $\delta A_F(t)$ are larger and can reach 15% but are still quite moderate.

## 3. Results and discussion

Figure 2 reports spatial variations of time- and transverse-averaged (a) fuel consumption and (b) heat release rates. It shows that the magnitude of the mean rate $\overline{\dot{\omega}_F}$ or $\overline{\dot{\omega}_T}$, sampled directly from the DNS data (see curves plotted in black solid lines), is overestimated if the counterpart filtered rate $\overline{\widehat{\dot{\omega}_F}}$ or $\overline{\widehat{\dot{\omega}_T}}$ is calculated using filtered density, temperature, and species mass fractions, i.e., $\widehat{\dot{\omega}_F} = \dot{\omega}_F(\hat{\rho}, \hat{T}, \hat{Y}_k)$ or $\widehat{\dot{\omega}_T} = \dot{\omega}_T(\hat{\rho}, \hat{T}, \hat{Y}_k)$, respectively, followed by time and transverse averaging (see curves plotted in color broken lines). Note that the rates $\overline{\dot{\omega}_F(\hat{\rho}, \hat{T}, Y_k)}$ and $\overline{\dot{\omega}_T(\hat{\rho}, \hat{T}, Y_k)}$ calculated by filtering the fields $\dot{\omega}_F(\mathbf{x},t)$ and $\dot{\omega}_T(\mathbf{x},t)$, respectively, followed by time- and transverse-averaging, are close [36] to the counterpart time- and transverse-averaged rates $\overline{\dot{\omega}_F}$ and $\overline{\dot{\omega}_T}$, respectively, sampled from the DNS data (not shown for brevity). In Fig. 2, differences between $\overline{\dot{\omega}_F}$ or $\overline{\dot{\omega}_T}$ (black solid lines) and the "no-combustion-model" (noCM) rate $\overline{\widehat{\dot{\omega}_F}}$ or $\overline{\widehat{\dot{\omega}_T}}$, respectively, are substantial even if the filter width $\Delta$ is as low as $0.44\delta_L^T$ (red dotted lines), with the differences being about 100% if $\Delta = 0.89\delta_L^T$ (violet dashed lines) or even much larger if $\Delta = 1.67\delta_L^T$ (blue dotted-dashed lines).

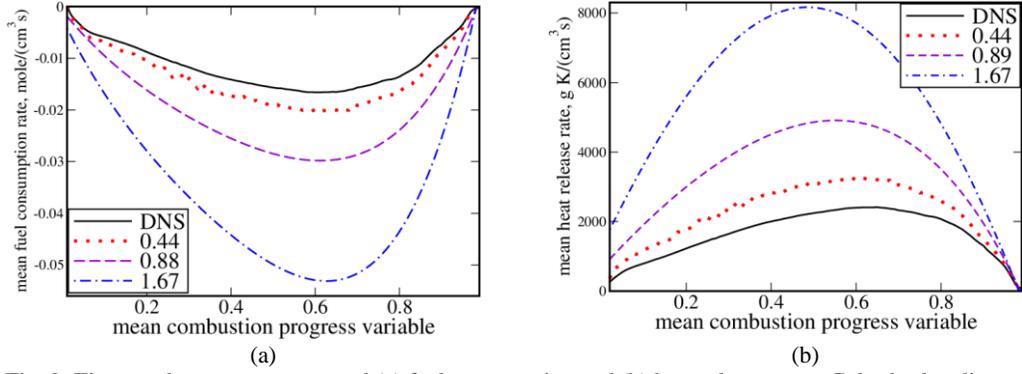

Fig. 2. Time- and transverse-averaged (a) fuel consumption and (b) heat release rates. Color broken lines show the rates evaluated using quantities filtered over various boxes whose normalized widths $\Delta/\delta_L$ are reported in legends. Black solid lines show the rates sampled from the DNS data.

On the contrary, Fig. 3 shows that differences in $\overline{|\nabla c_F|}$ or $\overline{|\nabla c_T|}$ sampled directly from the DNS data (black dots) and $\overline{|\nabla \hat{c}_F|}$ or $\overline{|\nabla \hat{c}_T|}$, respectively, are very small if $\Delta = 0.33\delta_L^T$ (red solid lines) and remain moderate even if the filter width $\Delta$ is larger than $\delta_L^T$ (orange dotted-double-dashed and violet double-dotted-dashed lines).

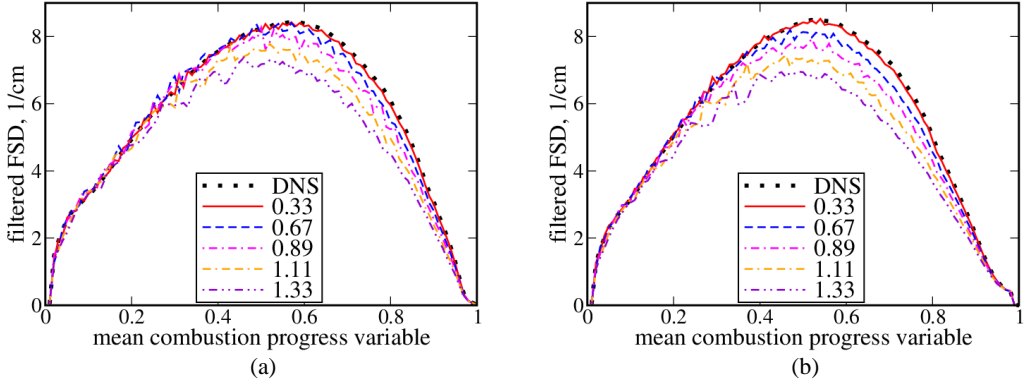

Fig. 3. Time- and transverse-averaged (a) fuel-based and (b) temperature-based generalized flame surface densities $\overline{|\nabla \hat{c}|}$ filtered (lines) using various boxes whose normalized widths $\Delta/\delta_L^T$ are reported in legends. Black dots show $\overline{|\nabla c|}$ sampled directly from the raw DNS data.

Thus, comparison of Figs. 2 and 3 implies that LES resolution requirements are significantly softer when invoking an FSD-based model of premixed turbulent burning when compared to direct calculation of filtered reaction rates using filtered density, temperature, and species mass fractions. For instance, Fig. 4a shows that comparable relative errors $\max\{|\bar{q}(\bar{c}) - \bar{\hat{q}}(\bar{c})|\}/\max\{|\bar{q}(\bar{c})|\}$ are obtained for $q = \dot{\omega}_F$ or $\dot{\omega}_T$ using $\Delta = 0.44\delta_L^T$ and for $q = |\nabla c_F|$ or $|\nabla c_T|$ using $\Delta = 1.33\delta_L^T$, see horizontal and vertical straight dotted-dashed lines. Here, $\bar{q}(\bar{c})$ and $\bar{\hat{q}}(\bar{c})$ designate time- and transverse-averaged DNS and filtered fields, $q(\mathbf{x}, t)$ and $\hat{q}(\mathbf{x}, t)$, respectively. It is worth stressing that (i) an increase in resolution by a factor of three (from $\Delta = 0.44\delta_L^T$ to $\Delta = 1.33\delta_L^T$) reduces numerical costs of unsteady three-dimensional simulations by a factor or $3^4 = 81$ and (ii) comparison of the influence of $\Delta$ on different quantities ($\bar{\dot{\omega}}_F$ or $\bar{\dot{\omega}}_T$ and $\overline{|\nabla c_F|}$ or $\overline{|\nabla c_T|}$, respectively) is justified in Fig. 1, which shows that almost the same turbulent burning velocity is obtained by integrating $\dot{\omega}_F(\mathbf{x}, t)$ or $\dot{\omega}_T(\mathbf{x}, t)$ and $S_L|\nabla c_F(\mathbf{x}, t)|$ or $S_L|\nabla c_T(\mathbf{x}, t)|$, respectively, where $S_L$ is constant.

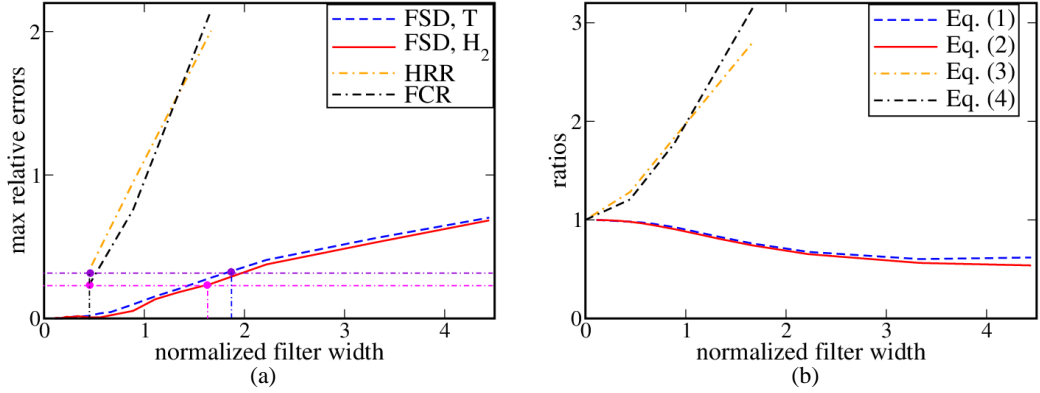

Fig. 4. (a) Maximal relative errors in evaluation of mean Flame Surface Density (FSD), Heat Release Rate (HRR), and Fuel Consumption Rate (FCR) yielded by filtered fields. (b) Ratio of time-averaged $\overline{U_t}$ (orange dotted-dashed and black dotted-double-dashed lines) or time-averaged $\overline{\delta A}$ (red solid and blue dashed lines) obtained from filtered fields to the counterpart quantity, $\overline{U_t}$ or $\overline{\delta A}$, respectively, sampled directly from DNS data.

The highlighted difference in resolution requirements is also observed in Fig. 4b, which reports ratios of (i) time-averaged turbulent burning velocity $\overline{U_t}$ (orange dotted-dashed or black dotted-double-dashed line) or (ii) time-averaged flame-surface-density integral $\overline{\delta A}$ (red solid or blue dashed line), obtained from filtered fields, to the counterpart quantity, $\overline{U_t}$ or $\overline{\delta A}$, respectively, sampled directly from the DNS data. The $\overline{U_t}$-ratios increase rapidly with $\Delta/\delta_L^T$ and are about three at $\Delta/\delta_L^T = 1.67$. On the contrary, the $\overline{\delta A}$-ratios decrease slowly with $\Delta/\delta_L^T$ and are higher than 0.5 even if $\Delta/\delta_L^T$ is as large as 4.44. Therefore, even if LES is performed adopting a coarse mesh, flame surface area can be predicted reasonably well without any model of subgrid flame wrinkling by small-scale turbulence, whereas a significantly finer mesh is required to reach the same level of prediction by directly evaluating fuel consumption and heat release rates without any combustion model, i.e., by calculating the rates using the filtered fields of $\hat{\rho}(\mathbf{x}, t), \hat{T}(\mathbf{x}, t)$, and $\hat{Y}_k(\mathbf{x}, t)$.

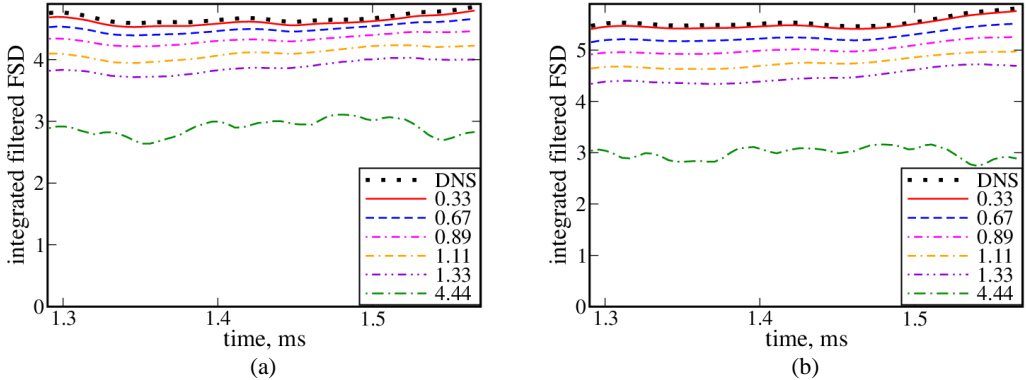

Fig. 5. Evolution of the axial integrals $\delta A(t)$ of (a) fuel-based and (b) temperature-based flame surface density filtered using various boxes whose normalized widths $\Delta/\delta_L^T$ are reported in legends.

Utility of the FSD-based noCM approach to LES of premixed turbulent flames is further demonstrated in Fig. 5, which reports (a) $\delta A_F(t)$ and (b) $\delta A_T(t)$, either sampled directly from the DNS data (black dots) or computed using various normalized filter widths $\Delta/\delta_L^T$ (color lines), specified in legends. Even if $\Delta/\delta_L^T$ is as large as 1.33, direct integration of $|\nabla \hat{c}_F|(\mathbf{x}, t)$ or $|\nabla \hat{c}_T|(\mathbf{x}, t)$ predicts $\delta A_F(t)$ or $\delta A_T(t)$, respectively, with a reasonable accuracy of 20-25%, cf. curves plotted in violet double-dotted-dashed lines and black dots. When $\Delta/\delta_L^T$ is decreased to 0.89, the errors are further reduced by a factor of about too, see curves plotted in magenta dotted-dashed lines.

It is worth stressing that this filter width is on the order of the Taylor length scale $\lambda = 0.69\delta_L$. Therefore, the filtered quantities are associated with the inertial range of Kolmogorov turbulence, whose characteristics are controlled by the mean dissipation rate, but are weakly affected by large-scale flow peculiarities [66]. Accordingly, confinement effects, which stem from a moderate ratio (13.3 in the present case) of computational domain width to $\delta_L^T$ and are typical for contemporary DNS studies of three-dimensional complex-chemistry turbulent flames characterized by $Ka > 1$, are not expected to change the present findings about resolution requirements.

Moreover, results of the present *a priori* study, reported in Figs. 1-5, imply that the source term $\overline{\rho\omega_c}(\mathbf{x}, t)$ in the well-known transport equation [67]

$$\frac{\partial}{\partial t}(\bar{\rho}\tilde{c}) + \nabla \cdot (\bar{\rho}\tilde{\mathbf{u}}\tilde{c}) = \nabla \cdot (\bar{\rho}\tilde{\mathbf{u}}\tilde{c} - \bar{\rho}\widehat{\mathbf{u}c}) + \overline{\rho\omega_c} \quad (5)$$

for the Favre-filtered combustion progress variable $\tilde{c}(\mathbf{x}, t) \equiv \widehat{\rho c}/\hat{c}$ can simply be evaluated as follows: $\overline{\rho\omega_c} = \rho_u S_L |\nabla\tilde{c}|$, i.e., without any model of subgrid-scale effects, because $\nabla\tilde{c}$ is directly computed using solution to Eq. (5). Such a noCM approach is expected to work reasonably well in equidiffusive mixtures provided that (i) filter width is comparable to thermal laminar flame thickness or less and (ii) local combustion quenching by intense turbulence does not play a statistically important role (low or moderately high Karlovitz numbers). Nevertheless, the approach does not resolve all issues, because subgrid turbulent transport [67] still requires modeling, see the first term on the right-hand side of Eq. (5), as well as thermal expansion effects reviewed elsewhere [68]. Therefore, while the presented results support the discussed noCM approach, they call also for its assessment in *a posteriori* LES studies.

It is worth noting that, while Figs. 3-5 report results obtained by directly filtering $c(\mathbf{x}, t)$-fields, analyses of the counterpart Favre-filtered fields $\tilde{c}(\mathbf{x}, t)$ yield similar results, e.g., see Fig. 6, which shows flame-surface areas computed by substituting $\hat{c}_T(\mathbf{x}, t)$ and $\hat{c}_F(\mathbf{x}, t)$ in Eqs. (1) and (2), respectively, with the Favre-filtered $\tilde{c}_T(\mathbf{x}, t)$ and $\tilde{c}_F(\mathbf{x}, t)$, respectively. The DNS data plotted in black dotted lines are the same in Figs. 5 and 6.

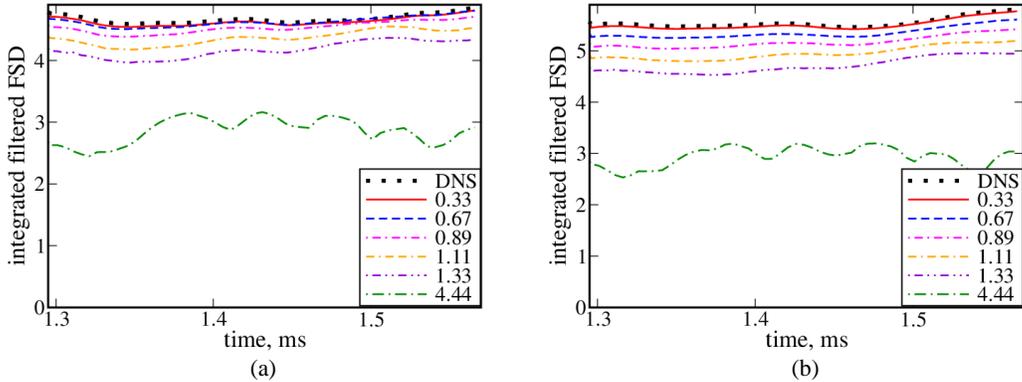

Fig. 6. Evolution of the axial integrals $\delta A(t)$ of (a) fuel-based and (b) temperature-based flame surface density Favre-filtered using various boxes whose normalized widths $\Delta/\delta_L^T$ are reported in legends.

It is also worth stressing that reasonable performance of noCM LES approach at $\Delta/\delta_L^T \approx 1$, shown in Figs. 3-5, does not result from weak fluctuations of $|\nabla c_F|(\mathbf{x}, t)$ or $|\nabla c_T|(\mathbf{x}, t)$ simulated by Dave et al. [53,54]. On the contrary, under conditions of that DNS study, such fluctuations are significant. For instance, Fig. 7 shows Probability Density Functions (PDFs) for normalized flame surface density $\delta_L^T|\nabla c_F|$ (violet dotted, black solid, and magenta dashed lines) and $\delta_L^T|\nabla c_T|$ (dotted-dashed lines) conditioned to the local values of $0.015 < c(\mathbf{x}, t) < 0.25$ (violet dotted and blue double-dotted-dashed lines), $0.045 < c(\mathbf{x}, t) < 0.55$ (black solid and brown dotted-dashed lines) and $0.075 < c(\mathbf{x}, t) < 0.85$ (magenta dashed and red dotted-dashed lines). These PDFs are wide, thus indicating significant fluctuations of $|\nabla c|(\mathbf{x}, t)$. Moreover, the PDFs peak at $\delta_L^T|\nabla c_T| \cong 1.2$ if $0.015 < c_T(\mathbf{x}, t) < 0.25$ or even $\delta_L^T|\nabla c_F| > 1.3$ if $0.015 < c_F(\mathbf{x}, t) < 0.25$. Such significant perturbations in the fields $|\nabla c|(\mathbf{x}, t)$ are associated with the influence of moderately large (when compared to $\delta_L^T$) turbulent eddies, which strain the local inherently laminar flame. On the contrary, the reasonable performance of noCM LES approach at $\Delta/\delta_L^T \approx 1$ is attributed to weak influence of small-scale (when compared to $\delta_L^T$) eddies on the local flame structure, see also Ref. [48].

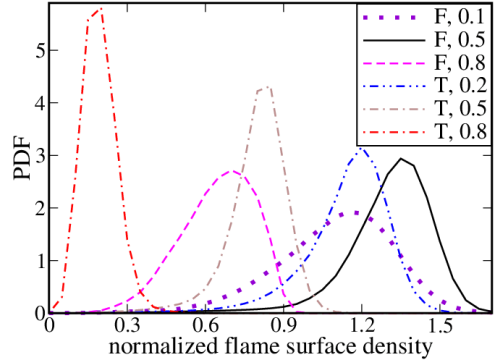

Fig. 7. Probability density functions for normalized flame surface density $\delta_L^T |\nabla c_F|$ (violet dotted, black solid, and magenta dashed lines) and $\delta_L^T |\nabla c_T|$ (dotted-dashed lines) conditioned to the local values of $c_F(\mathbf{x},t)$ and $c_T(\mathbf{x},t)$, respectively, specified in legends. The PDFs are sampled from the entire computational domain at 55 instants.

Finally, different resolution requirements to directly evaluating filtered reaction rates $\widehat{\dot\omega}$ and filtered flame surface density $|\widehat{\nabla c}|$ are associated with different challenges to be addressed in these two cases. First, small-scale phenomena affect both $\widehat{\dot\omega}$ and $|\widehat{\nabla c}|$ and should be resolved in both cases. However, experimental and numerical results overviewed briefly in Sect. 1, see Refs. [43-52] and papers quoted therein, imply that such phenomena could be properly resolved if $\Delta \approx \delta_L$.

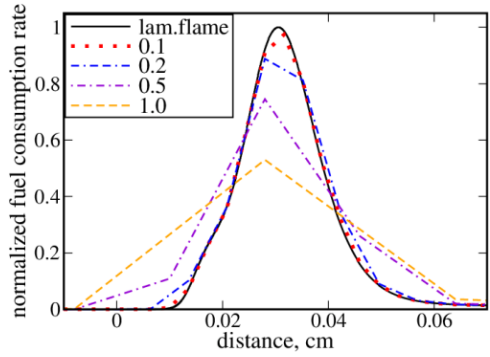

Fig. 8. Normalized fuel consumption rates obtained from laminar flame (black solid line) or computed using density, temperature, and species mass fractions filtered with a box whose normalized width $\Delta/\delta_L^T$ is reported in legends.

Second, filtering reaction rates involve an extra challenge resulting from highly non-linear dependencies of $\dot\omega_F$ and $\dot\omega_T$ on temperature [67]. Indeed, let us consider the simplest problem of a planar one-dimensional laminar flame propagating in quiescent mixture. Application of a box filter to such a flame yields the following well-known, purely numerical phenomenon [67], which does not result from any physical process. For instance, Fig. 8 compares a spatial profile of the normalized fuel consumption rate $\dot\omega_F(x)/\max\{\dot\omega_F(x)\}$ obtained from the laminar flame (black solid line) associated with the DNS by Dave et al. [53,54], with the rate $\widehat{\dot\omega_F}(x)/\max\{\dot\omega_F(x)\}$ calculated using $\hat\rho$, $\widehat{T}$, and $\widehat{Y_k}$ computed by applying different filters to the aforementioned laminar flame. If $\Delta/\delta_L^T = 0.1$, differences between $\dot\omega_F(x)$ and $\widehat{\dot\omega_F}(x)$ are small (red dots), but such differences become notable at $\Delta/\delta_L^T = 0.2$ (blue dotted-double-dashed line), significant at $\Delta/\delta_L^T = 0.5$ (violet dotted-dashed line), and large at $\Delta/\delta_L^T = 1.0$ (yellow dashed line). This simple well-known [67] example clearly shows that errors yielded by noCM LES of a premixed turbulent flame can be of purely mathematical nature (averaging of a highly non-linear function over insufficiently small volume), rather than be controlled by any physical mechanism.

## 4. Conclusions

The present analysis of DNS data obtained earlier by Dave et al. [53,54] from a moderately lean $H_2$/air flame under conditions of sufficiently intense small-scale turbulence ($Ka > 1$ and a ratio of laminar flame thickness to Kolmogorov length scale is about 20) shows that the mean flame surface density and area can be predicted with acceptable accuracy (i) using a sufficiently wide filter, e.g., $\Delta/\delta_L^T = 4/3$, and (ii) spatially integrating the generalized flame surface density $|\nabla \widehat{c_F}|(\mathbf{x}, t)$ or $|\nabla \widehat{c_T}|(\mathbf{x}, t)$, evaluated directly by processing the filtered scalar field $\widehat{c_F}(\mathbf{x}, t)$ or $\widehat{c_T}(\mathbf{x}, t)$, respectively. Such an approach does not require a model of the influence of subgrid turbulent eddies on flame surface density even if LES mesh is moderately coarse ($\Delta$ is on the order of thermal laminar flame thickness). On the contrary, when calculating fuel consumption or heat release rate using filtered density, temperature, and species mass fractions, a significantly finer (by a factor of three) mesh is required to reach a comparable prediction level. This better performance of the flame-surface-density-based noCM approach when compared to the reaction-rate-based noCM approach is associated with (i) inability of small-scale turbulent eddies to substantially wrinkle flame surface and (ii) highly non-linear dependence of fuel consumption or heat release rate on temperature.

**Declaration of competing interest**

The author declares that he has no known competing financial interests or personal relationships that could have appeared to influence the work reported in this paper.

**Acknowledgements**

The financial support by Swedish Research Council (grant 2023-04407) and Chalmers Area of Advance "Transport" (grant C 2021-0040) is gratefully acknowledged. The author is very grateful to Prof. S. Chaudhuri and Dr. H. Dave for sharing their DNS data.## References

[1] D. Veynante, L. Vervisch, Turbulent combustion modeling, Prog. Energy Combust. Sci. 28 (2002) 193-266.
[2] A.N. Lipatnikov, J. Chomiak, Turbulent flame speed and thickness: phenomenology, evaluation, and application in multi-dimensional simulations, Prog. Energy Combust. Sci. 28 (2002) 1-73.
[3] J. Janicka, A. Sadiki, Large eddy simulation of turbulent combustion systems, Proc. Combust. Inst. 30 (2005) 537-547.
[4] H. Pitsch, Large-eddy simulation of turbulent combustion, Annu. Rev. Fluid Mech. 38 (2006) 453-482.
[5] C.J. Rutland, Large-eddy simulations for internal combustion engines - a review, Int. J. Engine Res. 12 (2011) 421-451.
[6] L.Y.M. Gicquel, G. Staffelbach, T. Poinsot, Large eddy simulations of gaseous flames in gas turbine combustion chambers, Prog. Energy Combust. Sci. 38 (2012) 782-815.
[7] E. Hawkes, S.R. Cant, A flame surface density approach to large-eddy simulation of premixed turbulent combustion, Proc. Combust. Inst. 28 (2000) 51-58.
[8] C. Fureby, A fractal flame-wrinkling large eddy simulation model for premixed turbulent combustion, Proc. Combust. Inst. 30 (2005) 593-601.
[9] Z.M. Nikolaou, N. Swaminathan, Assessment of FSD and SDR closures for turbulent flames of alternative fuels, Flow Turbulence Combust. 101 (2018) 759-774.
[10] A.N. Lipatnikov, S. Nishiki, T. Hasegawa, A DNS assessment of linear relations between filtered reaction rate, flame surface density, and scalar dissipation rate in a weakly turbulent premixed flame, Combust. Theory Modell. 23 (2019) 245-260.